# Machine learning-guided construction of an analytic kinetic energy functional for orbital free density functional theory

**Sergei Manzhos[1], Johann Lüder[2], Pavlo Golub[3], and Manabu Ihara[1]**

[1] School of Materials and Chemical Technology, Institute of Science Tokyo, Ookayama 2-12-1, Meguro-ku, Tokyo, 152-8552, Japan
[2] Department of Materials and Optoelectronic Science, National Sun Yat-sen University, 80424, No. 70, Lien-Hai Road, Kaohsiung, Taiwan
[3] Department of Theoretical Chemistry, J. Heyrovský Institute of Physical Chemistry, Academy of Sciences of the Czech Republic, 3 Dolejškova 2155, Libeň, 182 00 Praha 8, Czechia

E-mail: Manzhos.s.aa@m.titech.ac.jp

## Abstract

Machine learning (ML) of kinetic energy functionals (KEF) for orbital-free density functional theory (OF-DFT) holds the promise of addressing an important bottleneck in large-scale ab initio materials modeling where sufficiently accurate analytic KEFs are lacking. However, ML models are not as easily handled as analytic expressions; they need to be provided in the form of algorithms and associated data. Here, we bridge the two approaches and construct an analytic expression for a KEF guided by interpretative machine learning of crystal cell-averaged kinetic energy densities ($\bar{\tau}$) of several hundred materials. A previously published dataset including multiple phases of 433 unary, binary, and ternary compounds containing Li, Al, Mg, Si, As, Ga, Sb, Na, Sn, P, and In was used for training, including data at the equilibrium geometry as well as strained structures. A hybrid Gaussian process regression – neural network (GPR-NN) method was used to understand the type of functional dependence of $\bar{\tau}$ on the features which contained cell-averaged terms of the 4th order gradient expansion and the product of the electron density and Kohn-Sham effective potential. Based on this analysis, an analytic model is constructed that can reproduce Kohn-Sham DFT energy-volume curves with sufficient accuracy (pronounced minima that are sufficiently close to the minima of the Kohn-Sham DFT-based curves and with sufficiently close curvatures) to enable structure optimizations and elastic response calculations.

## 1. Introduction

The development of kinetic energy functionals (KEF) for orbital-free (OF) density functional theory (DFT) that would be sufficiently accurate for use in applied modeling of most classes of materials remains an outstanding challenge. Computational materials modeling at the electronic structure level is dominated by Kohn-Sham (KS) DFT [1] whose CPU cost with system size scales formally near-cubically. This severely limits the range of applicability of DFT, whereby KS-DFT calculations are routinely doable on model systems not exceeding $10^2$ to $10^3$ atoms in size. This excludes a range of practically important phenomena that are intrinsically large-scale, such as microstructure-driven properties, biomolecular systems, disordered systems, low-concentration doping, etc. [2–4]. The current reliance on force fields for atomistic modeling of respective structures and phenomena is problematic, as force fields present only a crude approximation of the interatomic interactions landscape and cannot model electronic structure and electronic properties that are needed for mechanistic understanding of large-scale phenomena.

While other approaches to address the scaling issue exist such as order-N DFT [5–9] and semiempirical methods [10–16], they come at a cost of sometimes limiting approximations or parameterizations and at this point still do not address the issue of *routine* large-scale ab initio calculations in substance, as either scaling or the prefactor of the computational cost remain unfavorable. We stress the need for routine rather than one-off large-scale ab initio calculations, that are needed to screen multiple compositions and structures, perform long time-scale dynamics calculations, etc.

A method possessing the capability of routine (with calculations on systems of $10^5$ atoms doable on a desktop within hours) large-scale ab initio calculations *par excellence* is OF-DFT. OF-DFT corresponds to the original formulation of DFT [17] whereby the energy of a system of atoms (atomic cores and electrons) is expressed as a functional of electron density, $E = E[\rho(\boldsymbol{r})]$. In the absence of accurate functionals of the density, KS DFT [1,18] offered a practical way of computing $E[\rho(\boldsymbol{r})]$ via one-electron orbitals-solutions of the Kohn-Sham equation $\psi_i(\boldsymbol{r})$:

$$\begin{aligned} E[\rho(\boldsymbol{r})] = -\frac{1}{2}\sum_{i=1}^{N_{el}} \int \psi_i^*(\boldsymbol{r})\nabla^2\psi_i(\boldsymbol{r})d\boldsymbol{r} \\ + \frac{1}{2}\iint \frac{\rho(\boldsymbol{r}')\rho(\boldsymbol{r})}{|\boldsymbol{r}-\boldsymbol{r}'|}d\boldsymbol{r}d\boldsymbol{r}' \\ + \int V_{ion}(\boldsymbol{r})\rho(\boldsymbol{r})d\boldsymbol{r} + E_{XC}[\rho(\boldsymbol{r})] \\ \equiv E_{kin}[\{\psi_k(\boldsymbol{r})\}] + E_H[\rho(\boldsymbol{r})] + E_{ion}[\rho(\boldsymbol{r})] \\ + E_{XC}[\rho(\boldsymbol{r})] \\ \rho(\boldsymbol{r}) = \sum_{i=1}^{N_{el}}|\psi_i(\boldsymbol{r})|^2 dr \end{aligned} \tag{1}$$

Unless stated otherwise, we use atomic units. $N_{el}$ is the number of electron pairs (we neglected spin and partial occupancy without loss of generality of the present discussion). $E_H[\rho(\boldsymbol{r})]$ in the Hartree energy (due to Coulombic interactions within the electron density), $E_{ion}[\rho(\boldsymbol{r})]$ is the energy due to Coulombic interactions of the electron density and the Coulombic potential due to ionic cores and any additional external electric fields, and $E_{XC}[\rho(\boldsymbol{r})]$ is the exchange-correlation (XC) term that ensures that the electron density and the total energy computed from the orbitals match those of the system computed with the full many-body electronic wavefunction. The integrals are over the extent of the system (i.e. over the periodic cell for periodic systems or all space for non-periodic systems). $E_{kin}[\{\psi_k(\boldsymbol{r})\}]$ is the kinetic energy dependent on the entire set $\{\psi_k(\boldsymbol{r})\}$ of occupied orbitals. It is only $E_{kin}$ that requires the calculation of orbitals for eq. (1), all other terms can be computed directly as functionals of the density. The orbitals are also used for the calculation and analysis of band structure and excitations. The formally cubic scaling of DFT comes from the need to compute the orbitals (to solve the Kohn-Sham equation).

Over decades, a range of exchange-correlation functionals were developed providing accuracies of structures, bandstructures, excitation energies, etc. sufficiently high for their use in applied modelling (i.e. possessing predictive power) of all major classes of materials: organic and inorganic, semiconducting and metallic, periodic and non-periodic. It could be said that the key bottleneck has shifted from the accuracy of DFT to its ability to compute larger, and therefore more realistic, systems. OF-DFT addresses this bottleneck by using a kinetic energy functional (KEF) of the electron density computed without recourse to the orbitals,

$$E_{kin}[\rho(\boldsymbol{r})] = \int \tau(\boldsymbol{r})d\boldsymbol{r} \tag{2}$$

where $\tau(\boldsymbol{r})$ is kinetic energy density (KED). $E[\rho(\boldsymbol{r})]$ can then be computed with linear scaling and a small prefactor, and the bottleneck shifts to the accuracy of the KEF. One can rely on existing XC functionals and aim at a KEF reproducing KS kinetic energy or KS KED,

$$\begin{aligned} E_{kin}[\rho(\boldsymbol{r})] = \frac{1}{2}\sum_{i=1}^{N_{el}} \int \psi_i^*(\boldsymbol{r})\nabla^2\psi_i(\boldsymbol{r})d\boldsymbol{r} \equiv \int \tau_{KS}(\boldsymbol{r})d\boldsymbol{r} \\ = \int \tau_{KS}^+(\boldsymbol{r})d\boldsymbol{r} \end{aligned} \tag{3}$$

where $\tau_{KS}(\boldsymbol{r}) = \frac{1}{2}\sum_{i=1}^{N_{el}}\psi_i^*(\boldsymbol{r})\nabla^2\psi_i(\boldsymbol{r})$ and $\tau_{KS}^+(\boldsymbol{r}) = \frac{1}{2}\sum_{i=1}^{N_{el}}|\nabla\psi_i(\boldsymbol{r})|^2$ are Kohn-Sham KED and its positively-definite version (integrating to the same kinetic energy), respectively.

Existing analytic approximations for $E_{kin}[\rho(\boldsymbol{r})]$ already provide KS-DFT-like accuracy for light metals [19–26]. For other classes of materials, major issues remain, including the ability to correctly model potential energy surfaces - a prerequisite for structure optimization without which the method cannot be used in most applications. In recent years, there have been substantial developments enabling mechanistic analyses and calculations of optical spectra without recourse to orbitals [27–33]. Progress has also been made in the development of OF-DFT-suited pseudopotentials [34–37], including with the help of machine learning (ML) [22,38,39]. Several OF-DFT-capable codes are available where new KEFs can be eventually implemented [31,40–43]. The onus thus shifts to building good KEF models for different classes of materials including metals, semiconductors, and molecular materials. While steady progress is being made [30,44–49], improvements are needed. In the absence of a universal $E_{kin}[\rho(\boldsymbol{r})]$, widespread use of OF-DFT applications can succeed as sufficiently accurate $E_{kin}[\rho(\boldsymbol{r})]$ approximations are built for specific materials classes and or applications, similar to the state of KS DFT where a plethora of XC functionals are currently used for different applications.

Renewed interest in OF-DFT and specifically KEF research has been spurred in recent years by the possibility of machine-learning the $E_{kin}[\rho(\boldsymbol{r})]$ [50–64]. Most works, including those of the present authors [51–53,62], machine-learned KS KED from density-dependent features that typically include derivatives and powers of the electron density. When using the terms of the 4th order gradient expansion [65]

$$\tau_{GE4}(\boldsymbol{r}) = \tau_{TF}(\boldsymbol{r})\left(1 + \frac{5}{27}p(\boldsymbol{r}) + \frac{20}{9}q(\boldsymbol{r}) + \frac{8}{81}q(\boldsymbol{r})^2 - \frac{1}{9}p(\boldsymbol{r})q(\boldsymbol{r}) + \frac{8}{243}p(\boldsymbol{r})^2\right) \tag{4}$$

where $\tau_{TF}(\boldsymbol{r})$ is the Thomas-Fermi KED [66], and $p(\boldsymbol{r}) = \frac{|\nabla\rho(\boldsymbol{r})|^2}{4(3\pi^2)^{2/3}\rho(\boldsymbol{r})^{8/3}}$ and $q(\boldsymbol{r}) = \frac{\Delta\rho(\boldsymbol{r})}{4(3\pi^2)^{2/3}\rho(\boldsymbol{r})^{5/3}}$ are the scaled squared gradient and scaled Laplacian of the density, respectively, as well as the product of the electron density and Kohn-Sham effective potential $\rho(\boldsymbol{r})v_{eff}(\boldsymbol{r})$ as features, we were able to reproduce energy-volume curves of several materials simultaneously with an ML model based on kernel regression [52,62].

One disadvantage of ML of $\tau_{KS}(\boldsymbol{r})$ is the need to wield large data sets, with each structure of a material contributing possibly millions of data points. For example, sampling $\tau_{KS}(\boldsymbol{r})$ of Li, Al, and Si on their respective Fourier grids (of plane wave reference DFT calculations) resulted in about 600,000 datapoints [52,62]. This makes growing the training set to achieve a more portable KEF difficult, especially for materials with larger unit cells. Another challenge is the very uneven distribution of $\tau_{KS}(\boldsymbol{r})$ and of terms of eq. (4) which complicates ML [51,52,62]. In Ref. [62], this was addressed by smoothing $\tau_{KS}(\boldsymbol{r})$ and the features, which allowed accurate ML with relatively few training points. Fully averaging (over the simulation cell) would make each material structure contribute a single point to the training set. While this can inhibit machine learning from a smaller number of structures, when the number of materials in the training set is large enough, this is advantageous, as was done in Ref. [67] when machine-learning crystal cell-averaged KED ($\bar{\tau}$) of 433 unary, binary, and ternary compounds containing Li, Al, Mg, Si, As, Ga, Sb, Na, Sn, P, and In, including data at the equilibrium geometry as well as strained structures. Machine-learning cell-averaged KED rather than the integrated $E_{kin}$ is advantageous as it palliates the issue of different materials having different crystal cell sizes. In this work, we also machine learn $\bar{\tau}$ from the same data as in Ref. [67].

Another disadvantage of ML models of a KEF is that they are not as easily handled as analytic expressions can be. They need to be provided in the form of algorithms and associated data, as was done for example in Refs. [67,68]. This causes additional complexity, compared to an analytic formula, of integrating the model into various codes. Often the DFT engine (whether KS or OF) is written in a low-level language (such as Fortran or C++) while ML is done in Python frameworks like PyTorch, in Matlab, or other ML-directed environments. Complex ML models also inhibit analytic manipulations important for theory development. The black-box nature of generic ML methods such as neural networks, kernel regression, and gradient boosted decision trees to which the problem of machine learning a KEF is amenable also typically means lack of insight. Domain knowledge-specific ML methods are possible [69–72], but they lose one of the appeals of black-box ML methods which is precisely their generic nature.

Here, we build an analytic model of a kinetic energy functional, guided by a machine learning algorithm. We use the hybrid Gaussian process regression – neural network (GPR-NN) method [73] which, while fully general, allows for the analysis of the type of functional dependence of the target on the features, including the roles of nonlinearity and inter-feature coupling. GPR-NN was used previously for insightful ML in materials informatics and other applications [74–79]. Here, we use it to derive an analytic model of a KEF that is able to reproduce Kohn-Sham DFT energy-volume curves with good accuracy.

## 2. Methods and data

### *2.2 Data*

We use the dataset provided in Ref. [67] that includes the crystal cell-averaged KED $\bar{\tau}$ (the target) and crystal cell-averaged terms of the 4th order gradient expansion as well as crystal cell- averaged product of the electron density and Kohn-Sham effective potential, forming a six-dimensional feature vector $\boldsymbol{x} = \left(\overline{\tau_{TF}}, \overline{\tau_{TF}p}, \overline{\tau_{TF}p^2}, \overline{\tau_{TF}pq}, \overline{\tau_{TF}q^2}, \overline{\rho v_{eff}}\right)$. The term $\tau_{TF}q(\boldsymbol{r})$ is not used as it averages to zero. The data are computed with plane wave KS DFT and local pseudopotentials in Abinit [80], using the PBE functional [81]. It is computed for 433 unary, binary, and ternary compounds containing Li, Al, Mg, Si, As, Ga, Sb, Na, Sn, P, and In, including data at the equilibrium geometry as well as strained structures (18 structures for each composition, including the equilibrium structure as well as isotropically strained structures ranging from $0.93V_0$ to $1.10V_0$, where $V_0$ is the equilibrium volume), for a total of 7794 data points. These elements were chosen because local pseudopotentials (needed for OF-DFT) of a similar kind were available for them [22,23,38,82,83]. The materials were chosen as those available in Materialsproject database [84] with energies above hull not exceeding 400 meV. See Ref. [67] for other computational details. Pseudopotentials were used because reference KS-DFT calculations use plain waves and because a good kinetic energy model is more feasible when working with the smoother valence density. A KEF fitted to these data is therefore expected to have applicability for similar types of materials and calculations.

The features were scaled on the unit cube unless indicated otherwise. 80% of the data were used for training and 20% for testing unless otherwise indicated. We confirmed that there is no significant difference between different random training-test splits; we therefore fix the random seed in the following to facilitate comparisons.

### 2.2 Machine learning

We use the GPR-NN method of Ref. [73] which is effectively a hybrid between a single-hidden layer NN and GPR. A target function $f(\boldsymbol{x}), \boldsymbol{x} \in R^D$ (which in this case is $\bar{\tau}(\boldsymbol{x}), \boldsymbol{x} \in R^6$ ) is represented as a first-order additive model in redundant coordinates $\boldsymbol{y} \in R^N, N > D$:

$$f(\boldsymbol{x}) = \sum_{n=1}^{N} f_n(y_n) \tag{5}$$

where the component functions $f_n(y_n)$ are built with GPR

$$f_n(y_n) = \sum_{k=1}^{M} a_{nk} k\left(y_n, y_n^{(k)}\right) \tag{6}$$

where $k\left(y_n, y_n^{(k)}\right)$ is the kernel and $k$ indexes training data points. The kernel is one-dimensional and therefore issues with high-dimensional kernels [79,85,86] are avoided. We use the square exponential kernel, i.e. $k(y, y') = \exp((y - y')^2/2l^2)$. The redundant coordinates $y_n$ are linear functions of $\boldsymbol{x}$, $\boldsymbol{y} = \boldsymbol{W}\boldsymbol{x}$ , where $\boldsymbol{W}$ is defined by rules and is not optimized. Here, the original coordinates $\boldsymbol{x}$ are included as a subset of $\boldsymbol{y}$ ($y_n = x_n$ for $n = 1, ..., D$) and the rows of matrix $\boldsymbol{W}$ defining other $y_n$ are chosen as elements of a $D$-dimensional Sobol sequence [87]. All terms $f_n(y_n)$ of eq. (5) are computed as a single linear step with a standard GPR code where one defines an additive kernel in $\boldsymbol{y}$ [88,89]. The only computational cost overhead vs. standard GPR is the summation in the kernel. Functions $f_n(y_n)$ are optimal in the least squares sense for given $\boldsymbol{W}$ and data.

Eq. (5) is equivalent to a single-hidden layer NN with $N$ neurons, a linear output neuron, and optimal shapes of neuron activation functions for each neuron. It possesses therefore a universal approximator property [90–104]. Note that biases and output weights are subsumed in the definition of $f_n(y_n)$ and need not be considered separately. As no nonlinear optimization is done, the method is as robust as linear regression (as GPR is a regularized linear regression with nonlinear basis functions derived from the kernel function [79,105]), and there is no overfitting as $N$ exceed the optimal number of neurons [73].

Control of $N$ allows studying importance of features, functional dependency of the target on features, and importance of coupling [74,75,79]. Here, we use this capability of the method to build an analytic model of a KEF based on machine learning results.

Calculations are performed in Matlab R2022b. The code for the GPR-NN method is provided in Ref. [73]. The version of the code modified for the present application can be obtained from the authors. Hyperparameters are selected by grid search. Optimal values were around $l$ = 0.5 for the length parameter of the (isotropic in $\boldsymbol{y}$) kernel and $\log \sigma$ = -5 for the logarithm of the noise parameter. While there are slight variations in the optimal values with $N$, they are not significant, so that the hyperparameters were fixed at these values.

The quality of energy-volume curves was estimated by the error in the quantity characterizing the curvature of the energy volume curve at its minimum,

$$B = V_0^2 \frac{d^2E}{dV^2} \approx B' = \frac{E(V_0 - \Delta V) - 2E(V_0) + E(V_0 + \Delta V)}{(\Delta V/V_0)^2} \tag{7}$$

where $V_0$ is the equilibrium volume, $\Delta V/V_0$ = 3%, and $E$ the total energy. The total energy with the ML KEF model was estimated by subtracting KS kinetic energy from KS DFT total energy and replacing it with model-predicted kinetic energy (which is the product of $\bar{\tau}$ and the volume).

## 3. Results

We first fitted $\bar{\tau}(\boldsymbol{x})$ with GPR-NN for different $N$ to understand obtainable accuracy with the method vs. previous calculations on the data. In Ref. [67], the best GPR model of that work obtained an RMSE (root mean square error) in $\bar{\tau}$ of $1.05\times10^{-5}$ and $1.47\times10^{-5}$ a.u. for the training and test sets, respectively. This corresponded in the error in $B'$ of 13% in terms of MRE (mean absolute relative error) and 8% in terms of MDRE (median absolute relative error), and energy-volume curves of the vast majority of the 433 materials sufficiently close to KS DFT computed curves to result in meaningful structure optimization, stiffness calculations, etc. To achieve such accuracy, correlation coefficients $R$ between reference and model-predicted $\bar{\tau}$ had to exceed 0.99999 ("five nines"). To compare, linear regression resulted in RMSE in $\bar{\tau}$ of about $4\times10^{-5}$ a.u. (for both the training and test sets), MRE and MDRE of about 55% and 25%, respectively, and energy-volume curves of many materials that did not exhibit a minimum around the equilibrium geometry. This even though $R$ of about 0.9999 ("four nines") was high by the standards of most ML applications. This highlights that ML for KEF development required extremely accurate ML. This is of course not surprising as interaction energies defining energy-structure dependence are small compared to the total energy.

In Table 1, we summarize accuracy achievable with GPR-NN with different numbers of terms $N$. Note that $N$ = 6

corresponds to a simple additive model in $\boldsymbol{x}$ ($\boldsymbol{y} = \boldsymbol{x}$). Key observations from these results are

(i) GPR-NN achieves an accuracy competitive with full-dimensional GPR already with an additive model. While levels of RMSE in $\bar{\tau}$ are a tad higher than with GPR, the accuracy of *B'* is similar and is slightly better. The correlation coefficient approaching "five nines" is obtained.

(ii) increasing *N* i.e. adding coupling terms, further decreases RMSE in $\bar{\tau}$ to better levels than was achieved with GPR, *R* is also increased beyond "five nines", but with limited additional benefit to the quality of the energy volume curves exemplified by MRE and MDRE of *B'* values. This is not surprising, as the ML algorithm only aims to improve the aggregate error in $\bar{\tau}$, while the errors in *B'*, that are essentially interaction energy errors, are influenced by any error cancellations or accumulation.

We therefore take a closer look at the additive model corresponding to *N* = 6 ("6" in the table). This is the model whose component functions we will fit with polynomials. We found that slightly better low-order polynomial fits can be obtained if we manually adjust the scaling of some coordinates (thus introducing an effect that an anisotropic kernel would have while keeping the kernel isotropic). The results are given in the table under "6' ". The fit quality is slightly worse but allows obtaining slightly better low-order polynomial fits introduced below (where the scaling factors are those defining $\boldsymbol{x}'$ in Eq. (9)).

Table 1. Errors in $\bar{\tau}$ and in *B*' achieved with GPR-NN with different numbers of terms *N*. *N* = 6 corresponds to first-order additive model in $\boldsymbol{x}$. Also given is the correlation coefficient *R* between Kohn-Sham and predicted $\bar{\tau}$ values.

| | Error in $\bar{\tau}$, $\times 10^5$ *a.u.* | | Error in *B'*, % | | *R*, digits xxx in 0.9999xxx | |
|---|---|---|---|---|---|---|
| *N* | training | test | MRE | MDRE | train | test |
| 6 | 1.81 | 1.87 | 10.5 | 6.4 | 863 | 844 |
| *6'* | *1.87* | *1.91* | *10.9* | *7.5* | *853* | *837* |
| 20 | 1.35 | 1.38 | 10.3 | 6.8 | 924 | 915 |
| 50 | 1.21 | 1.23 | 10.6 | 8.0 | 939 | 932 |
| 100 | 1.15 | 1.16 | 10.6 | 7.9 | 945 | 939 |

Figure 1 shows the correlation plot between the exact (Kohn-Sham DFT-based) and ML model-predicted $\bar{\tau}$ from the additive model ($\boldsymbol{y} = \boldsymbol{x}$) for the model corresponding to "6' " in Table 1 as well as the importance of component functions by the square root of their variance. The correlation plot is practically linear; we show it here just to highlight the level of accuracy required in this application vs mainstream ML applications such as e.g. materials informatics or ML force fields. The plot of relative importance of features indicates the following:

(i) A large fraction of the variance of the target is captured by $\overline{\rho v_{eff}}$. This is expected, because the KS KED can be expressed as $\tau_{KS}(\boldsymbol{r}) = \sum_i \epsilon_i |\psi_i(\boldsymbol{r})|^2 - v_{eff}(\boldsymbol{r})\rho(\boldsymbol{r})$, where $\epsilon_i$ are Kohn-Sham eigenvalues.

(ii) All features have important contributions. Indeed, we confirmed by doing calculations with some features removed from the set that a good accuracy cannot be obtained unless all of them are included.

(iii) We can compare relative feature importance vs. the gradient expansion. The leading (linear) polynomial coefficients at features $x_1, \dots, x_5$ allow a relatively direct comparison when feature $x_6$ not present in the gradient expansion (albeit responsible for a large fraction of the variance as per Figure 1) is excluded. For ease of comparison with Eq. (4), the result can be written as (note that the Laplacian term is not included as it averages to zero)

$$
\begin{aligned}
&\tau_{TF}(\boldsymbol{r})(1 + 0.162p(\boldsymbol{r}) + 0.062q(\boldsymbol{r})^2 - 0.364p(\boldsymbol{r})q(\boldsymbol{r}) \\
&\quad - 0.730p(\boldsymbol{r})^2) \\
&\quad = \tau_{TF}(\boldsymbol{r})\left(1 + \frac{4.44}{27}p(\boldsymbol{r}) + \frac{5.03}{81}q(\boldsymbol{r})^2 \right. \\
&\quad \left. - \frac{3.28}{9}p(\boldsymbol{r})q(\boldsymbol{r}) + \frac{-1.77}{243}p(\boldsymbol{r})^2\right)
\end{aligned} \tag{8}
$$

We observe that the 2nd order term follows closely the gradient expansion, while the other terms' leading polynomial coefficients deviate from it but are of the same order of magnitude. Importantly, the gradient expansion is *linear* in $x_1, \dots, x_5$, while the ML and the polynomial model are nonlinear and are fitted to reproduce the KE of a specific set of materials based on specific reference calculations, so that there is no requirement or expectation for the resulting coefficients to match closely (note that in often-studied TF+λvW functionals empirical λ values are often used rather than the gradient expansion value [30]). The exclusion of $x_6$ (i.e. $\overline{\rho v_{eff}}$) leads to a much less accurate model, with training/test rmse of about 3.7 / 3.9×10$^{-5}$ a.u.

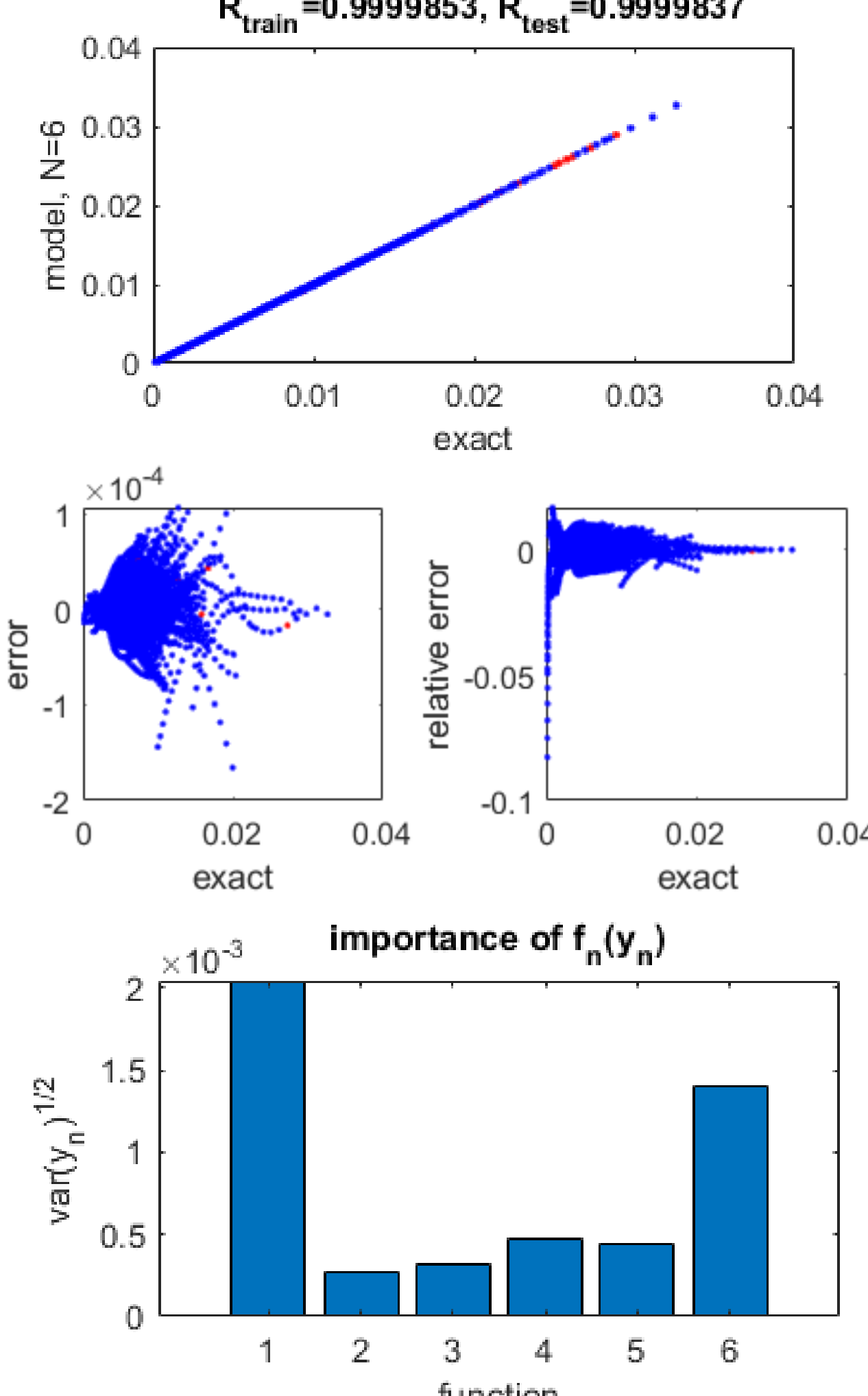

Figure 1. Top panel: the correlation plot between the exact (Kohn-Sham DFT-based) and model-predicted crystal cell-averaged kinetic energy density $\bar{\tau}$ from the additive model ($\boldsymbol{y} = \boldsymbol{x}$). The model corresponds to "6' " in Table 1. Blue points are from the training and red points from the test set (many are invisible due to the high quality of the fit). Corresponding correlation coefficients are shown on top of the plot. Middle panels: absolute and relative errors in $\bar{\tau}$. Bottom panel: the importance of component functions by the square root of their variance Numbers on the abscissa index $n$.

When including $x_6$(model *6'* in Table 1), we observe that the model assigns a weight (coefficient at the linear term in the polynomial) to the von Weizsäcker- term of about 0.133, which is somewhat smaller than gradient expansion's weight but is of the same order of magnitude. The relative magnitudes of the $\overline{\tau_{TF}p^2}$, $\overline{\tau_{TF}pq}$, $\overline{\tau_{TF}q^2}$ terms vs the TF term (about 0.152, 0.233, and 0.214, respectively) are somewhat larger than in the gradient expansion. The fact that the ML model returns physically meaningful relative magnitudes of the terms related to the terms of the gradient expansion is of note. This kind of analysis is not easily accessible with standard NN and kernel regressions and highlights the utility of GPR-NN in providing interpretative, insightful ML.

Nonlinearity is critically important: we can force the model to be linear by using a very large kernel length parameter [75,85]. For example, with $l$ = 1000, $\log \sigma = -10$ (when the component functions are practically linear), we obtain values of MRE and MDRE in $B$' of about 38 and 21%, respectively, and RMSE in $\bar{\tau}$ of about 4×10$^{-5}$ a.u. (comparable to results obtained in Ref. [67] with linear regression).

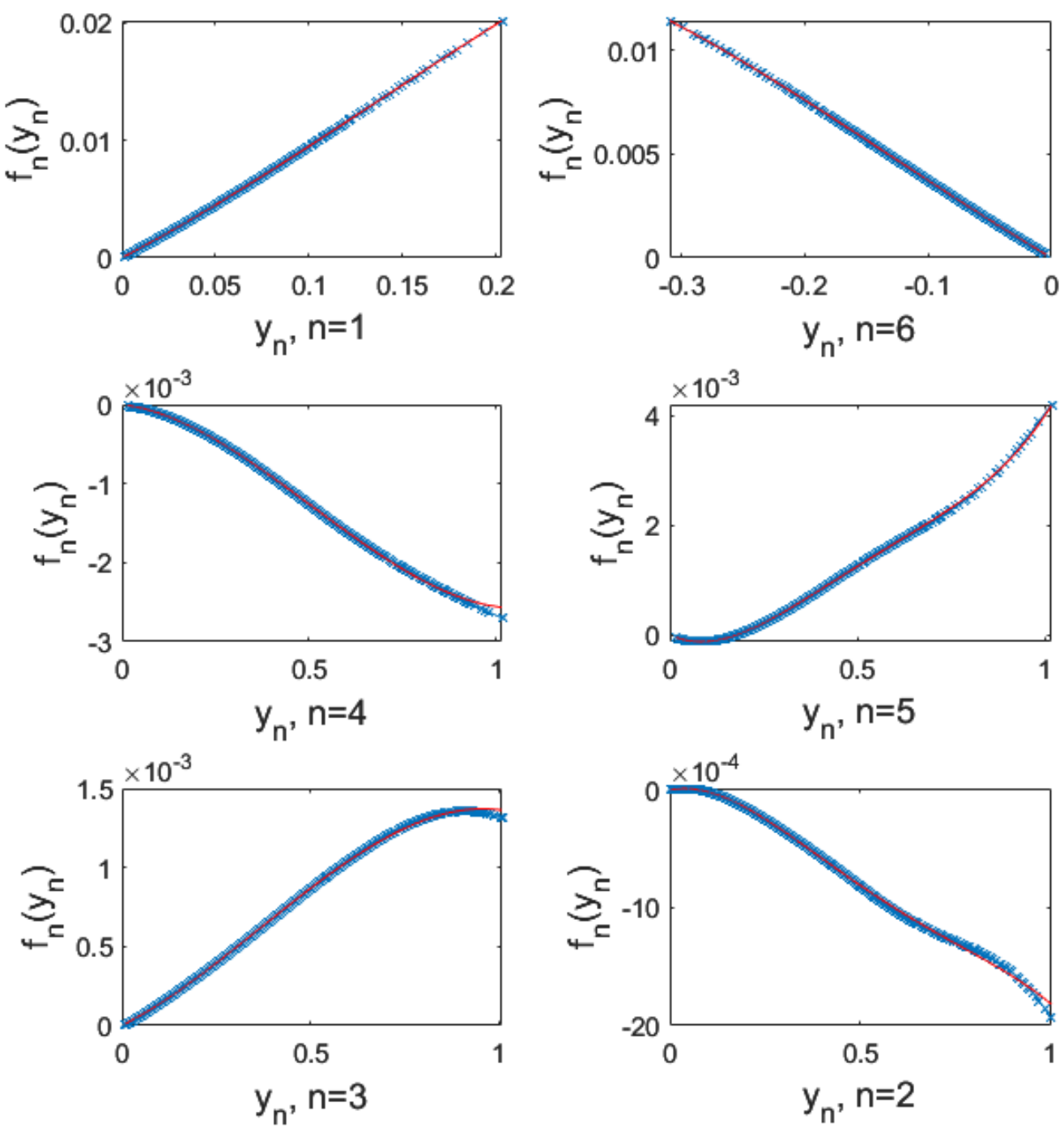

Figure 2. The shapes of the component functions $f_n(y_n)$, in the order of importance (left to right and top to bottom) predicted by the additive model ($\boldsymbol{y} = \boldsymbol{x}$). The symbols are the values at the training data points and the red line is the polynomial fit.

Figure 2 shows the shapes of the component functions. One can appreciate their nonlinear nature. On the other hand, they are sufficiently smooth to be potentially expressible with simple analytic functions.

We therefore fit them with low order polynomials (using all data points). The red line in Figure 2 shows the polynomial fit with 3$^{rd}$ order polynomials for $x_1, x_3, x_4$ and 4$^{th}$ order polynomials for $x_2, x_5, x_6$. Using lower-order polynomials resulted in significant deterioration of the accuracy. The energy volume curves are computed based on the polynomial model. The best polynomial model is obtained with slightly different hyperparameters ($l$ = 0.35, $\log \sigma = -4$). The polynomial model obtained RMSE in $\bar{\tau}$ of 2.21×10$^{-5}$ a.u. and MRE and MDRE in $B$' of 13.6% and 9.5%, respectively.

These values are somewhat worse than those achievable with the GPR-NN models but they are by and large similar to those achieved in Ref. [67] with GPR, except that now we can express the KEF as an analytical formula that can be communicated and handled as such. That formula is:

$$\bar{\tau} = \sum_{n=1}^{6} \sum_{p=1}^{P_n} a_{np} (x'_n)^p \tag{9}$$

where $\boldsymbol{P} = (3,4,3,3,4,4)$, $\boldsymbol{x}' = (6.6\overline{\tau_{TF}},\ 271\overline{\tau_{TF}p},\ 797\overline{\tau_{TF}p^2}, 655\overline{\tau_{TF}pq},\ 194\overline{\tau_{TF}q^2},\ 8.4\overline{\rho v_{eff}}\ )$, and matrix $\boldsymbol{a}$ with elements $a_{np}$ is

$$\begin{pmatrix} 0.082926586827897 & 0.157677447591406 & -0.378760872909400 & 0 \\ 0.000634147249392 & -0.009076865906296 & 0.011656254814405 & -0.005025491981521 \\ 0.001196170711360 & 0.001986140118683 & -0.001816638030194 & 0 \\ -0.000570356598613 & -0.005784298540582 & 0.003797894548784 & 0 \\ -0.003230090903822 & 0.023009315178709 & -0.030436676933612 & 0.014654972864815 \\ -0.035134852665972 & 0.027869731148558 & 0.062744904623325 & -0.029165231906284 \end{pmatrix} \tag{10}$$

In Figure 3 - Figure 6, we show the energy-volume curves computed with KS DFT (blue lines) and the analytic model (red lines) for the entire set of 433 materials (in the same order as they are provided in the dataset of Ref. [67] i.e. by increasing materialsproject (mp) ID which is also shown on the plots). On each graphic, we indicate, besides the chemical formula, the values of *B'* from KS DFT and model-predicted curves, in respective colors. For the vast majority of the materials, the ML-inspired analytic model results in pronounced minima that are sufficiently close to the minima of the KS DFT curves, and with sufficiently close curvatures to perform structure optimizations and elastic response calculations.

Overall, the results are not worse than those of a full GPR model (see energy-volume curves of Ref. [67]). Some of the few materials with unfavorable energy volume curves already were so in the GPR model of Ref. [67]. These cases can be seen in, for instance, cubic phases of P and As including mp-10 and mp-53, mp-7245 or the layered AsP (mp-1228795) structure. Those are materials with covalent bonds but in geometries quite different from their ground state, i.e., phases with relatively high energies above hull. Since the calculations generating the database were performed with DFT and local pseudopotentials, a contributing factor could be a limited transferability of the local pseudopotentials of elements like As and P forming covalent bonds. Hexagonal packed Si (mp-34), or tetragonal phases containing As or P such as mp-130, mp-8882 (GaP) or mp-570744 ($Si_3As_4$) further support this. Indeed, the ratio of obtaining well matched *B*' values is, for instance, about 2 times higher for phases containing Mg or 10 times for Sn than for P or Si. Difficulties of the model with layered systems arise from the averaging step that similarly affects structures with void regions such as mp-1072544, a Si phase. Another set of struggling energy volume curves can be seen in Na and Li containing materials where the DFT and GPR-NN curves have noticeable shifts, which could stem from the lack of (semi-) core electrons in the local pseudopotential [106]. Essentially, the ability of the ML model to reproduce the physics with energy volume curves of a broad variety of materials seems to be limited by the restrictions in the data stemming from the local pseudopotential KS DFT calculations, emphasizing the need for comprehensive approaches to simultaneously develop KEF and local pseudopotentials, or rely on all-electron approaches.

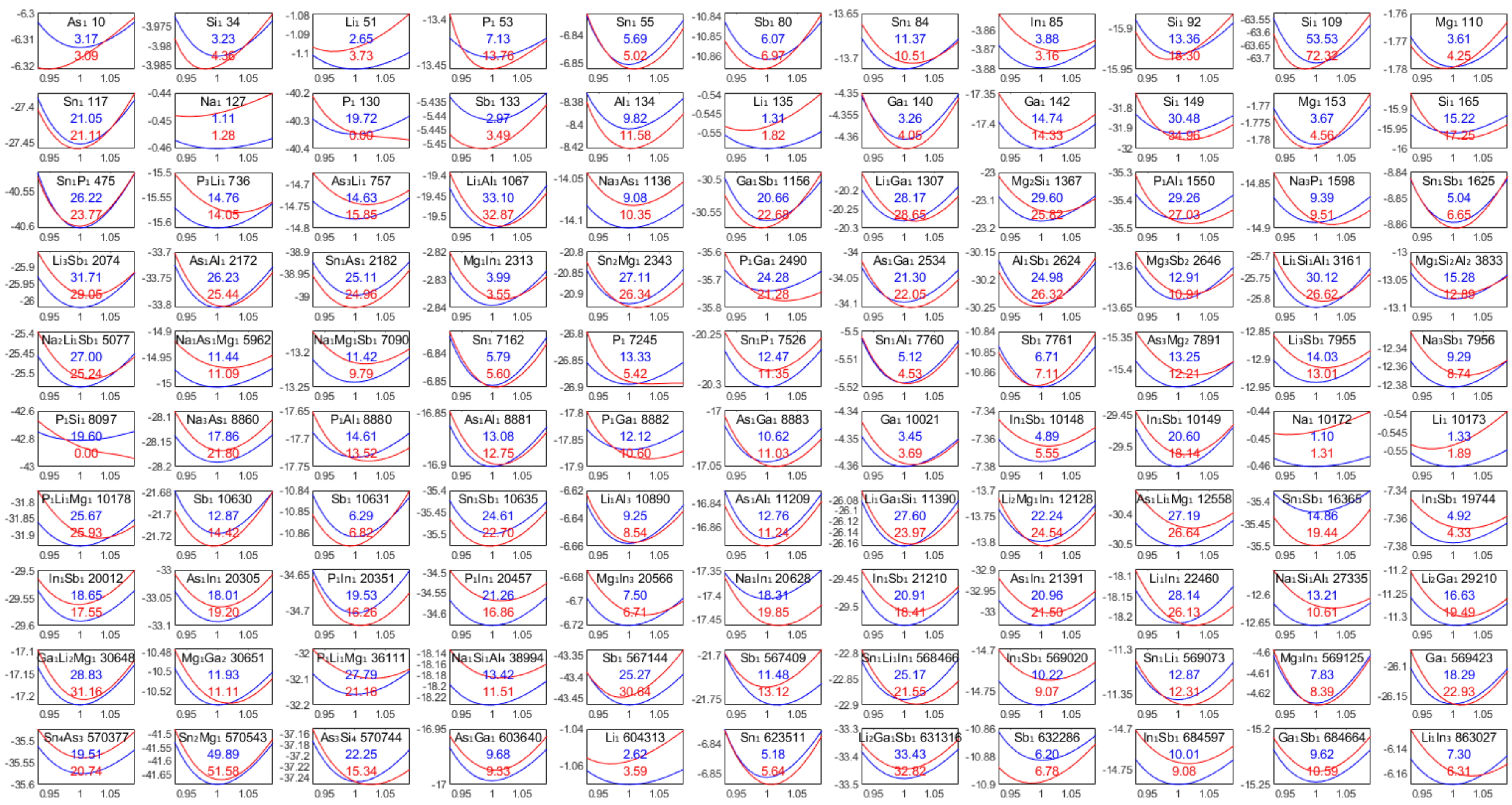

Figure 3. Energy-volume curves for the materials in the dataset. Chemical formula is followed by materialsproject ID. Blue: Kohn-Sham DFT; red: analytic model of $\bar{\tau}$. The abscissa values are $V/V_0$ with the equilibrium volume at $V/V_0 = 1$, and the ordinate axis is total energy in a.u. The values of $B'$ from KS DFT and model-predicted curves are shown blue and red, respectively.

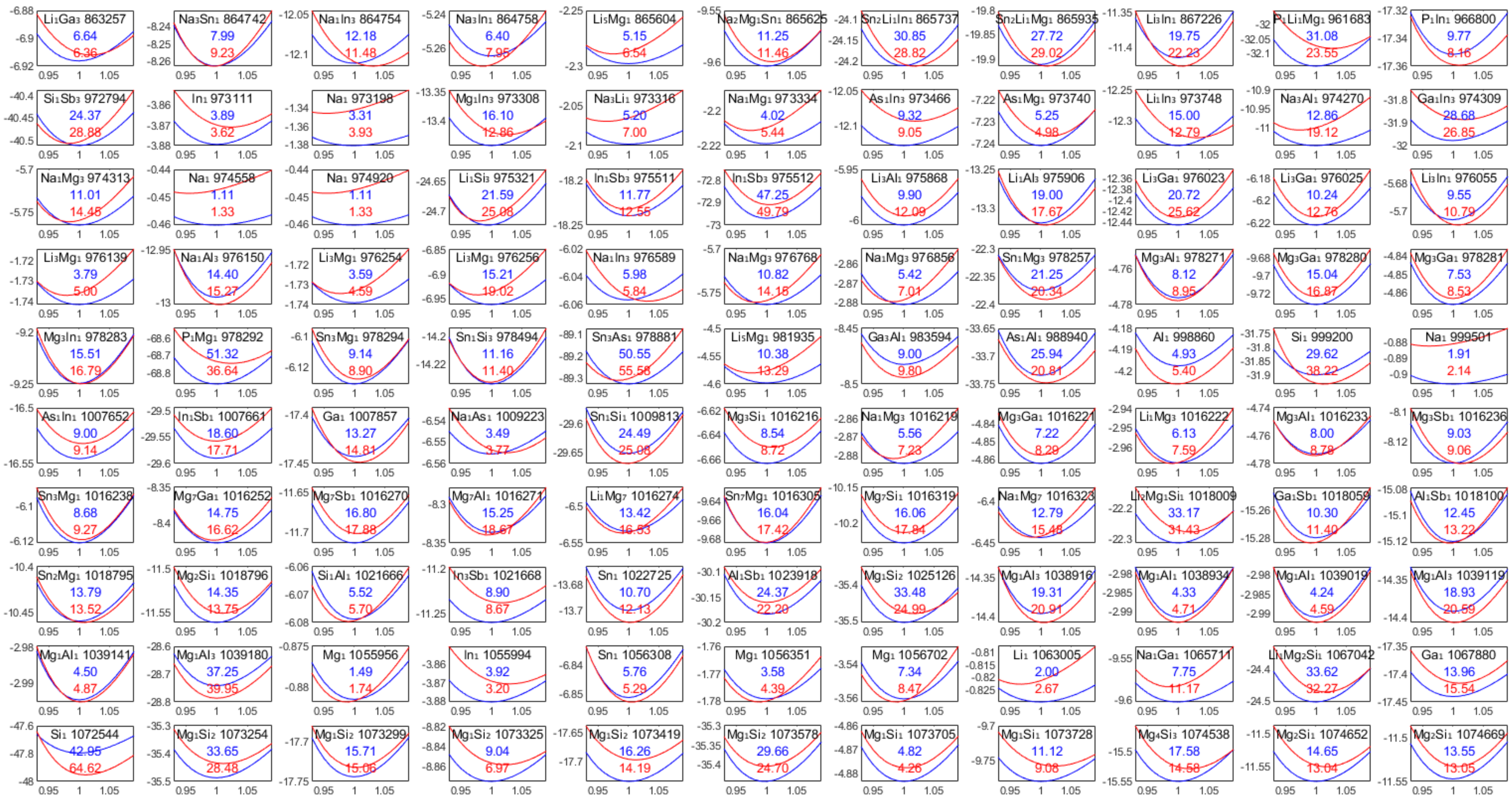

Figure 4. Energy-volume curves for the materials in the dataset (continued). Chemical formula is followed by materialsproject ID. Blue: Kohn-Sham DFT; red: analytic model of $\bar{\tau}$. The abscissa values are $V/V_0$ with the equilibrium volume at $V/V_0 = 1$, and the ordinate axis is total energy in a.u. The values of $B'$ from KS DFT and model-predicted curves are shown blue and red, respectively.

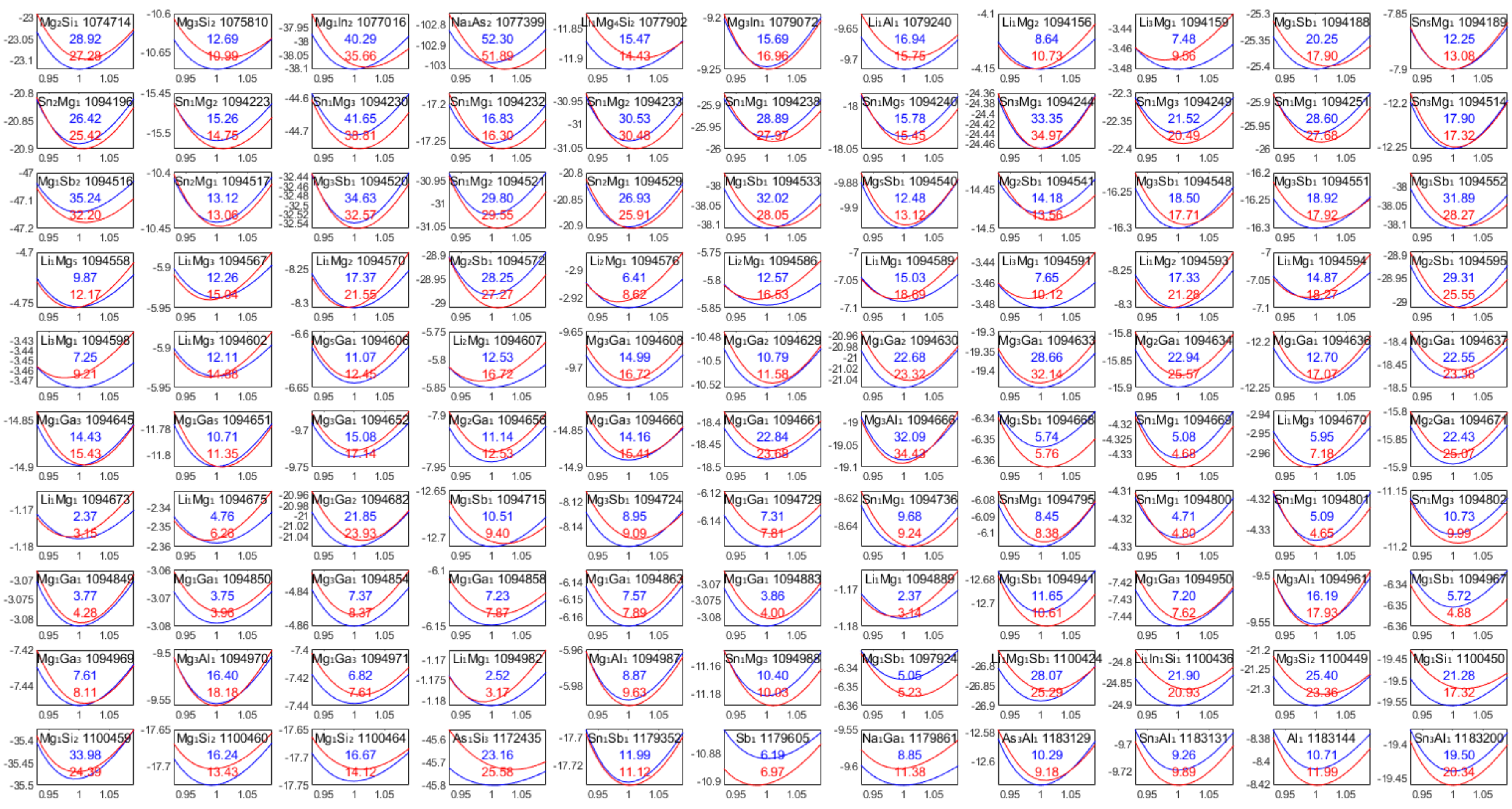

Figure 5. Energy-volume curves for the materials in the dataset (continued). Chemical formula is followed by materialsproject ID. Blue: Kohn-Sham DFT; red: analytic model of $\bar{\tau}$. The abscissa values are $V/V_0$ with the equilibrium volume at $V/V_0 = 1$, and the ordinate axis is total energy in a.u. The values of $B'$ from KS DFT and model-predicted curves are shown blue and red, respectively.

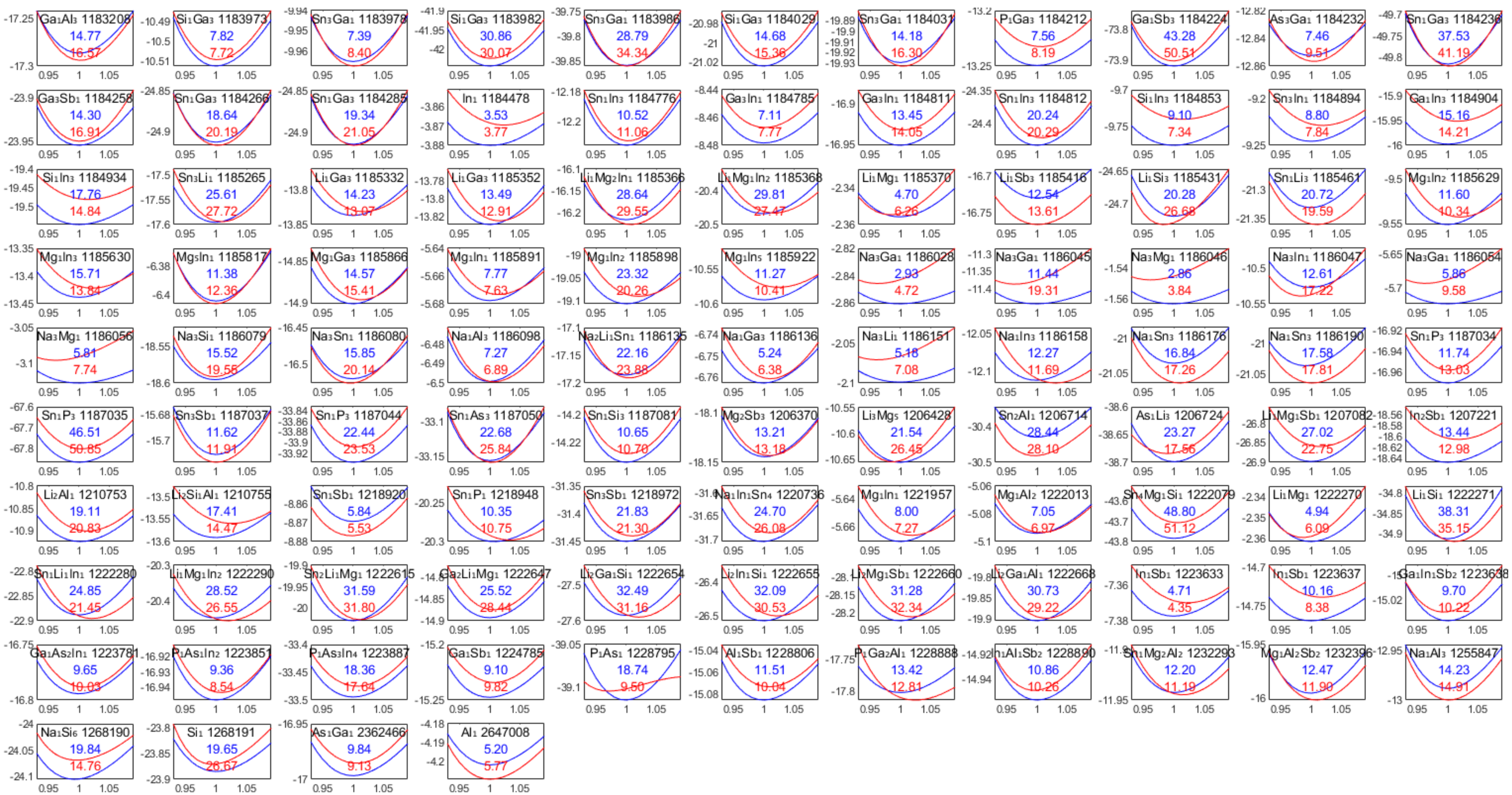

Figure 6. Energy-volume curves for the materials in the dataset (continued). Chemical formula is followed by materialsproject ID. Blue: Kohn-Sham DFT; red: analytic model of $\bar{\tau}$. The abscissa values are $V/V_0$ with the equilibrium volume at $V/V_0 = 1$, and the ordinate axis is total energy in a.u. The values of $B'$ from KS DFT and model-predicted curves are shown blue and red, respectively.

In principle, any ML model (NN, GPR, or GPR-NN) is expressible analytically (see eqs. (5-6), for example), but such expressions are unwieldy. In the case of GPR, for example, the number of summands is equal to the training data set size; and additional summation in the additive kernel further complicates the expression [107]. In the case of a NN, one sums over multiple layers, multiple neurons per layer, and over terms in the argument of neuron activation functions [108,109]. Such expressions cannot be manipulated as analytic formulas in most practically relevant applications. Equations (9-10) can be.

## 4. Conclusions

We used machine learning of crystal cell-averaged kinetic energy density to guide the construction of an analytic expression for a kinetic energy functional. We applied the GPR-NN method to the previously published dataset of 433 433 unary, binary, and ternary compounds containing Li, Al, Mg, Si, As, Ga, Sb, Na, Sn, P, and In [67]. Terms of the 4$^{th}$ order gradient expansion and the product of the electron density and Kohn-Sham effective potential that were used as ML features. GPR-NN, which is a hybrid between neural network and GPR, was able to produce fits of similar or better accuracy than previously reported GPR fits of the same data. Importantly, the use GPR-NN allowed insight into the relative importance of nonlinearity and inter-feature coupling and the shape of functional dependence of the target on the features. The quality of the ML model was evaluated by the quality of the energy-volume curves and the curvature at the minimum of the curve, thus directly addressing the issue of reproducing the topology of the interaction energy which remains a major issue in OF-DFT.

Based on the analysis of GPR-NN results, we constructed a relatively simple analytic expression that is practically as accurate as the best ML model. For most materials, it can reproduce Kohn-Sham DFT energy-volume curves with sufficient accuracy (pronounced minima that are sufficiently close to the minima of the Kohn-Sham DFT-based curves and with sufficiently close curvatures) to enable structure optimizations and elastic response calculations.

ML KEF models are actively pursued for their promise to facilitate large-scale ab initio materials modeling with OF-DFT, but they cannot be as easily manipulated (shared, used, analyzed) as analytic formulas - they have to be shared as codes and datasets. Our results effectively bridge the two tracks of KEF development. We hope that they will be informative for future ML applications including OF-DFT functional development for electronic problems and beyond [110], XC functional development [56,111–115] and other ML applications broadly.

The analytic model developed here is mimicking and ML model trained on a particular dataset. A KEF fitted to these data is therefore expected to have applicability for similar types of materials and calculations. There is still a lack of datasets and databases from which kinetic energy functionals for OF-DFT could be machine-learned. The availability of the dataset of Ref. [67] enabled the present work. More extensive datasets with more diverse compositions and structures as well as non-equilibrium densities are desired to build more accurate and general ML KEFs and ML-based analytic KEFs for which the scheme proposed here can be used.

One challenge of the model presented here is that the use of the terms of the 4$^{th}$ order gradient expansion means an OF-DFT code implementing such functional must have a robust density optimizer. We have tested the functional in the development version of CONUNDrum [53] and confirmed that it leads to stable optimization when initiated with Abinit densities (i.e. the functional did not “walk away” from the true densities) as well as stable optimization when initiated with WT [116] densities and to lower than WT energies, as shown in Figure 7 on the examples of fcc Al and cubic diamond Si, where examples of convergence of self-consistent density optimization are also shown. On the other hand the optimizer tended to get stuck when initialized with very non-equilibrium densities, which was traced to difficulties of a simple optimizer (CONUNDrum uses a gradient descent based optimizer) dealing with numerically sensitive terms dependent on $\overline{\tau_{TF}p^2}$, $\overline{\tau_{TF}pq}$, $\overline{\tau_{TF}q^2}$, (i.e. higher derivatives and powers of the density with the density in the denominator). Nevertheless, these tests show that the functional works, and initialization with a less accurate existing functional like WT used here is a viable option. The development of datasets with non-equilibrium densities, and development of optimizers comfortably working with terms of the gradient expression are therefore important directions of future research in light of the present results.

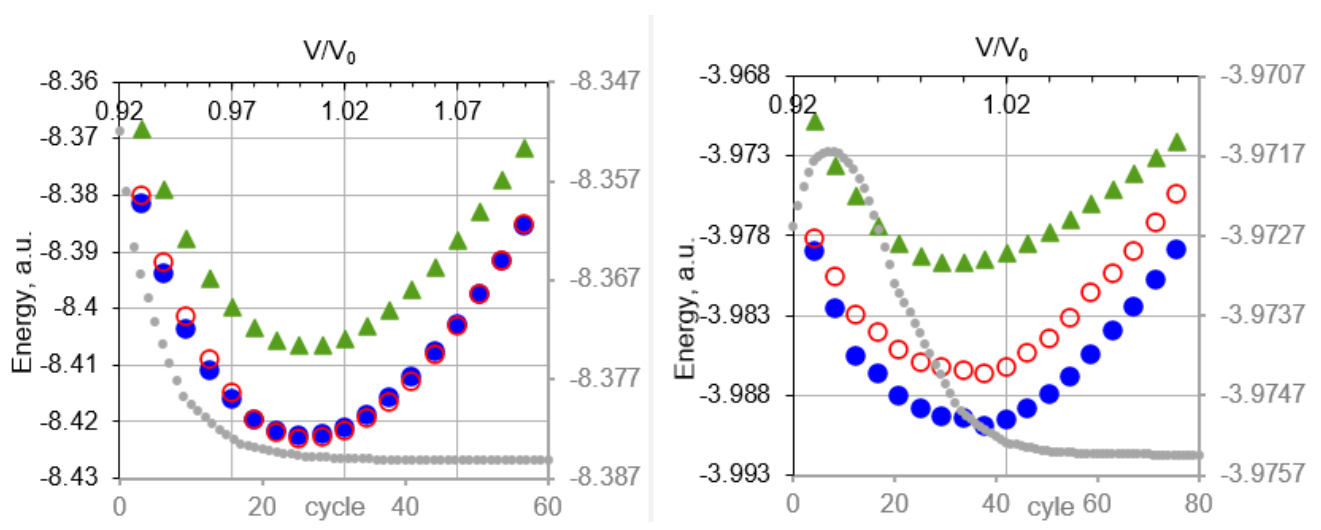

Figure 7. Energy-volume curves (corresponding to the black axes) for fcc Al (left) and cubic diamond Si (right). Filled blue circles: initialized with Abinit densities and optimized with the new KEF, empty red circles: initialized with WT densities and optimized with the new KEF, green triangles: WT energies before optimization with the new functional. The grey dots (corresponding to the grey secondary axes) illustrate the convergence of self-consistent density optimization (for $V/V_0 = 1.1$, WT initialization).

## Data availability statement

The data used in the manuscript are provided in Ref. [67]. The code for the GPR-NN method is provided in Ref. [73]. A version of the code used specifically in this manuscript is available in Supporting Information and is also available from the authors.

## Acknowledgements

S. M. and M. I. thank JST Mirai Program (grant No. JPMJMI22H1).

J. L. thanks the National Science and Technology Council (NSTC) for financial support under Grant N. 110-2112-M-110-025, and the National Center for High-performance Computing (NCHC) of National Applied Research Laboratories (NARLabs) in Taiwan for providing computational and storage resources.

## Author contributions

S. M.: conceptualization (lead); methodology (lead); software (lead); investigation (lead); writing - original draft (lead); writing - review and editing (lead).

J. L.: data curation (lead); investigation (supporting); writing - review and editing (supporting).

M. I: writing - review and editing (supporting); resources.

P. G.: investigation (supporting); writing - review and editing (supporting).

## Conflicts of interest

The authors declare no competing interests.